\documentclass{llncs}

\usepackage{algo}  
\usepackage{epsfig}  
\usepackage{url}  
\usepackage{latexsym}   
\usepackage{boxedminipage}  
\begin{document}

\newcounter{algo}  
\setcounter{algo}{0}  
\newenvironment{algo}[1]{%
\refstepcounter{algo}  
\begin{center}\sl Algorithm \thealgo: #1  
\begin{boxedminipage}{0.95\textwidth}  
}{\end{boxedminipage}\end{center}}

\newcommand{\mC}{\mathcal{C}}  
\newcommand{\mF}{\mathcal{F}}  

\newenvironment{preuve}{\noindent {\it Proof.}}{$\Box$\vskip1ex}  
\newenvironment{preuveA1}{\noindent {\it Proof of conservation of   
Property $A_1$.}}{$\Box$\vskip1ex}  
\newenvironment{preuveA2}{\noindent {\it Proof of conservation of   
Property $A_2$.}}{$\Box$\vskip1ex}  
\newenvironment{preuveB}{\noindent {\it Proof of the invariants.}}  
{$\Box$\vskip1ex}  
\newenvironment{OJO}{\noindent {\bf OJO:}}{$\Box$\vskip1ex}  
  
  
\title{A Note On Computing Set Overlap Classes}

\author{Pierre Charbit\inst{1}\quad
Michel Habib\inst{1} \quad
Vincent Limouzy\inst{1}\\
Fabien de Montgolfier\inst{1}\quad
Mathieu Raffinot\thanks{Corresponding author. E-mail: {\tt raffinot@liafa.jussieu.fr}}\inst{1}\quad Micha\"el Rao\inst{2}}

\institute{LIAFA, Univ. Paris Diderot - Paris 7, 75205 Paris Cedex 13, France.
\and
LIRMM, 161 rue Ada, 34392 Montpellier, France.}
\maketitle

\begin{abstract}
Let ${\cal V}$ be a finite set of $n$ elements and ${\cal F}=\{X_1,X_2,
\ldots , X_m\}$ a family of $m$ subsets of ${\cal V}.$ Two sets
$X_i$ and $X_j$ of ${\cal F}$ overlap if $X_i \cap X_j \neq
\emptyset,$ $X_j \setminus X_i \neq \emptyset,$ and $X_i \setminus X_j
\neq \emptyset.$ Two sets $X,Y\in {\cal F}$ are in the same overlap
class if there is a series $X=X_1,X_2, \ldots, X_k=Y$ of sets of
${\cal F}$ in which each $X_iX_{i+1}$ overlaps. In this note, we focus
on efficiently identifying all overlap classes in $O(n+\sum_{i=1}^m
|X_i|)$ time. We thus revisit the clever algorithm of Dahlhaus
\cite{Dahlhaus00} of which we give a clear presentation and that we
simplify to make it practical and implementable in its real worst case
complexity. An useful variant of Dahlhaus's approach is also explained.  
\end{abstract}  
  
\section{Introduction}  
\label{sec:intro}  

Let ${\cal V}$ be a finite set of $n=|{\cal V}|$ elements and ${\cal
  F}=\{X_1,X_2, \ldots , X_m\}$ a family of $m$ subsets of ${\cal V}.$
Two sets $X_i$ and $X_j$ of ${\cal F}$ overlap if $X_i \cap X_j \neq
\emptyset,$ $X_i \setminus X_j \neq \emptyset,$ and $X_j \setminus X_i
\neq \emptyset.$ We denote $|{\cal F}|$ as the sum of the sizes of all
$X_i \in {\cal F}$. We define the overlap graph $OG({\cal F},E)$ as the
graph with all $X_i$ as vertices and $E=\{(i,j)\; |\; X_i \mbox{
  overlaps } X_j\}, \forall \; 1 \leq i,j \leq m.$ A connected
component of this graph is called an {\em overlap class.}

In this note we focus on efficiently identifying all overlap classes of
$OG({\cal F},E).$ This problem is a classical one in graph clustering
related topics but it also appears frequently in many graph problems
related to graph decomposition \cite{Dahlhaus00} or PQ-tree
manipulation \cite{McConnell04}.

An efficient $O(n+|{\cal F}|)$ time algorithm has already been
presented by Dahlhaus in \cite{Dahlhaus00}. The algorithm is very
clever but uses an off-line Lowest Common Ancestor algorithm (LCA) as
subroutine. From a theoretical point of view, off-line LCA queries
have been proved to be solvable in constant time (after a linear time
preprocessing) in a RAM model (accepting an additional constant time
specific register operation) but also recently in a pointer machine
model \cite{BKRW98}. However, in practice, it is very difficult to
implement these LCA algorithms in their real linear
complexity. Another difficulty with Dahlhaus's algorithm comes from
that its original presentation is difficult to follow. These two
points motivated this note. Dahlhaus's algorithm is really clever and
deserves a clear presentation, all the more so we show how to replace
LCA queries by set partitioning, which makes Dahlhaus's algorithm
easily implementable in practice in its real complexity. We also
provide a source code freely available in \cite{OurImpl07}. We
eventually explain how to simply modify Dahlhaus's approach to
efficiently compute a spanning tree of each connected component of the
overlap graph. This simplifies a graph construction in
\cite{McConnell04}.

\section{Dahlhaus's algorithm}
\label{dahlhausalgorithm}

The overlap graph $OG({\cal F},E)$ might have $\Theta(m^2)$ edges,
which can be quadratic in $O(|{\cal F}|).$ For instance, if ${\cal F}
=\{\{x_1,x_2\}, \{x_1,x_3\}, \ldots, \{x_1,x_m\}\}$,
$|E|=m(m-1)/2=\Theta(m^2).$

The approach of Dahlhaus is quite surprising since that, instead of
computing a subgraph of the overlap graph, Dahlhaus considers
a second graph $D({\cal F},L)$ on the same vertex set but with
different edges. This graph has however a strong property: its
connected components are the same than that of $OG({\cal F},E)$,
although that in the general case $D({\cal F},L)$ is not a subgraph
of $OG({\cal F},E).$
 
Let $\mbox{LF}$ be the list of all $X \in {\cal F}$ sorted in
decreasing size order. The ordering of sets of equal size is arbitrarily
fixed. Given $X \in {\cal F}$, we denote $\mbox{Max}(X)$ as the largest
$Y \in {\cal F}$ taken in $LF$ order such that $|Y| \geq |X|$ and $Y$
overlaps $X$. Note that $\mbox{Max}(X)$ might be undefined for some
sets of ${\cal F}.$ In this latter case, in order to simplify the
presentation of some technical points, we write
$\mbox{Max}(X)=\emptyset.$ Dahlhaus's algorithm is based on the
following observation:

\begin{lemma}[\cite{Dahlhaus00}]
\label{thelemma}
Let $X \in {\cal F}$ such that $\mbox{Max}(X)\not=\emptyset$. Then for
all $Y \in {\cal F}$ such that $Y \cap X \neq \emptyset$ and $ |X|
\leq |Y| \leq |\mbox{Max}(X)|$, $Y$ overlaps $X$ or $\mbox{Max}(X).$
\end{lemma}
\begin{preuve}
If $Y$ does not overlap $X$, as $|X| \leq |Y|$ and $Y \cap X \neq
\emptyset$, $X \subseteq Y.$ Thus $Y \cap \mbox{Max}(X) \neq
\emptyset.$ Then, if $Y$ does not overlap $\mbox{Max}(X)$, then
$\mbox{Max}(X) \subseteq Y$. But in this case, as $|Y| \leq
|\mbox{Max}(X)|,$ $Y=\mbox{Max}(X)$ and overlaps $X$. Therefore $Y$
overlaps $X$ or $\mbox{Max}(X).$
\end{preuve}

Let us assume that we already computed all $\mbox{Max}(X).$ For each
$v \in {\cal V}$ we compute the list $SL(v)$ of all sets $X \in {\cal
  F}$ to which $v$ belongs. This list is sorted in increasing order of
the sizes of the sets. Computing and sorting all lists for all $v\in
{\cal V}$ can be done in $O(|{\cal F}|)$ time using a global bucket
sort.

Dahlhaus's graph $D({\cal F},L)$ is built on those lists. Let $X$ be a set
containing $v$ such that $\mbox{Max}(X)\not=\emptyset$. Then for all
consecutive pairs $YW$ after $X$ in $SL(v)$ ($X$ included, i.e. $Y$
can be instanced by $X$) and such that $|W|\leq |\mbox{Max}(X)|$,
create an edge $(Y,W)$ in the graph $D$.

\begin{lemma}[\cite{Dahlhaus00}]
The two graphs $D({\cal F},L)$ and $OG({\cal F},E)$ have the same
connected components.
\label{samecomponent}
\end{lemma}
\begin{preuve}
$(\Rightarrow)$ Let $Y,W \in {\cal F}$ such that $(Y,W) \in
L.$ By construction there exists $v$ such that $Y$ and $W$ are
consecutive on $SL(v)$ and there exists $X$ that appears before
$YW$ on $SL(v)$ such that $\mbox{Max}(X)\not=\emptyset$ and such that
$|X|\leq |Y| \leq |W| \leq |\mbox{Max}(X)|.$ By lemma \ref{thelemma},
$Y$ and $W$ overlap either $X$ or $\mbox{Max}(X).$ As $X$ and
$\mbox{Max}(X)$ overlap, the sets $X$, $Y$, $W$, and $\mbox{Max}(X)$ belong to
the same overlap class of $OG({\cal F},E)$. By extension, the vertices
of any connected path in $D({\cal F},L)$ belong to the same overlap
class of $OG({\cal F},E)$. 

\noindent
$(\Leftarrow)$ Let $A, B \in {\cal F}$ be two overlapping sets, {\em
  i.e. } $(A,B) \in E.$ Let $v \in A \cap B.$ Assume w.l.o.g. that
  $|A|\leq |B|.$ Then $\mbox{Max}(A)\not=\emptyset$ and $|\mbox{Max}(A)|
  \geq |B|.$ Therefore, in $SL(v)$, there exits a serie of consecutive
  pairs $YW$ from $A$ to $B$ that are linked in $D({\cal F},L).$ In
  consequence, $A$ and $B$ are connected in $D({\cal F},L).$
\end{preuve}

\begin{figure}[htb]
\vspace{-0.3cm}
  \centering
\includegraphics[width=8cm]{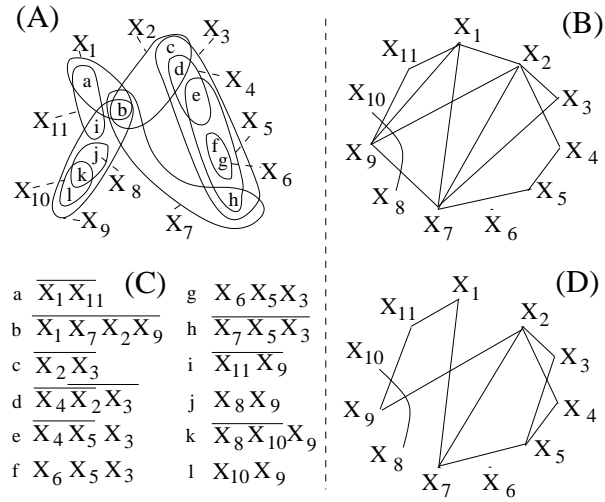}
\caption{Global example: (A) input family of 11 sets; (B) Overlap graph; (C) $SL$ lists; (D) Dahlhaus's graph. On (C) intervals defined by $\mbox{Max}(X)$ are overlined. Notice that Dahlhaus's graph is not a subgraph of the Overlap graph.}
 \label{allexample}
\vspace{-0.3cm}
\end{figure}

Notice that the order of equally sized sets in $SL$ lists has no
importance for the construction of a Dahlhaus's graph. Figure
\ref{allexample} shows an example of an overlap graph and a Dahlhaus's
graph.

\begin{lemma}[\cite{Dahlhaus00}]
Given all $\mbox{Max}(X), X \in {\cal F}$, the graph $D({\cal F},L)$
can be built in $O(|{\cal F}|)$ time and its number of edges is less
than or equal to $|{\cal F}|.$
\end{lemma}
\begin{preuve}
To build the graph $D({\cal F},L)$ from the $SL$ lists, it suffices to
go through each $SL$ list from the smallest set to the largest and
remenber at each step the largest $\mbox{Max}(X)$ already seen. If the size
of the current set is smaller than or equal to this value, an edge is
created between the last two sets considered.
  
Let us now consider the number of edges of $D({\cal F},L).$ As at most
one edge is created for each set in a list $SL$, at most $|{\cal F}|$
edges are created after processing all lists.
\end{preuve}

Identifying the overlap classes of $OG({\cal F},E)$ can therefore be
done by a simple Depth First Search on $D({\cal F},L)$ in $O(n+|{\cal
  F}|)$ time. It remains however to explain how to efficiently compute
all $\mbox{Max}(X).$

\section{Computing all $\mbox{Max}(X)$}

Let $\mbox{LF}$ be the list of all $X \in {\cal F}$ sorted in
decreasing size order. The order of sets of equal size is not
important. We consider a boolean matrix $\mbox{BM}$ of size $|{\cal
F}|\times |V|$ such that each row represents a set $X\in {\cal F}$
in the order of $\mbox{LF}$, and each column an element $v \in V.$ The
value $\mbox{BM}[i,j]$ is $1$ if and only if $v_j \in X_i.$

\begin{figure}[htb]
\vspace{-0.3cm}
  \centering
\includegraphics[width=7cm]{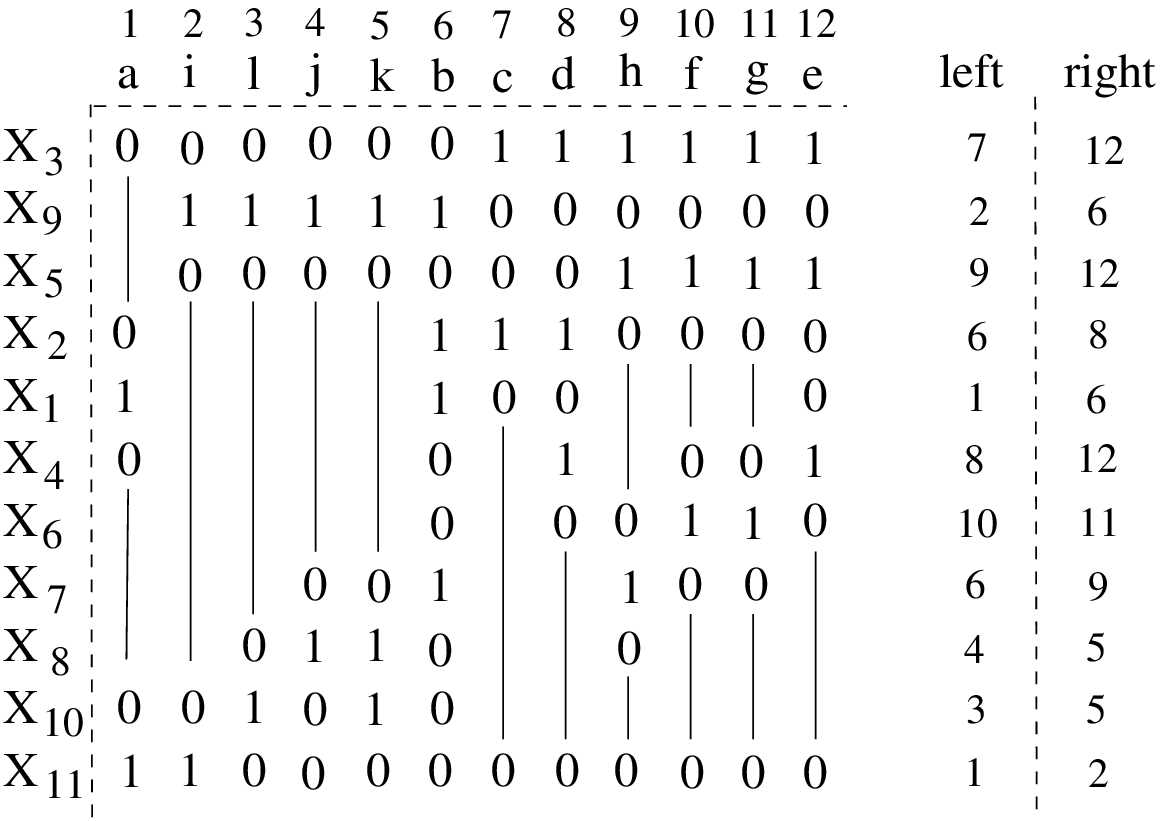}
\caption{Example continued: $BM$ matrix which lines are sorted by
decreasing sizes of $X\in {\cal F}$ and which columns are sorted in lexicographic order.}
 \label{matrix}
\vspace{-0.3cm}
\end{figure}

The first step of Dahlhaus's algorithm is to sort the columns of
$\mbox{BM}$ in lexicographical order, although that there is no detail
in \cite{Dahlhaus00} on how to do it efficiently in $O(|{\cal F}|)$
time. We postpone all explanations concerning this step to section
\ref{partitions} and we consider below that all columns of $BM$ are
lexicographically sorted. Figure \ref{matrix} shows the $BM$ matrix
for the set family of Figure \ref{allexample}.

For each $X \in {\cal F}$ we denote $\mbox{left}(X)$
(resp. $\mbox{right}(X)$) the number of the column of $BM$ containing the
leftmost (resp. rightmost) $1$ in the row of $X$.

\begin{lemma}
 Let $X,Y\in {\cal F}$ such that $Y$ overlaps $X$ and let $r_Y$ be the
 row of $Y$ in $BM$. Then there exists a row $t$ higher than or equal
 to $r_Y$ such that $\mbox{BM}[t,\mbox{left}(X)]=0$ and
 $\mbox{BM}[t,\mbox{right}(X)]=1.$
\label{goodlemma}
\end{lemma}
\begin{preuve}
 As $Y$ overlaps $X$, $|X|\geq 2.$ Let $r_X$ be the row
 corresponding to $X$ in $\mbox{BM}.$ Since $Y$ overlaps $X$, there
 exist two indices $1 \leq i < j \leq |V|$ and a row $r$ such that
 $\mbox{BM}[r_X,i]=\mbox{BM}[r_X,j]=1$, such that one of the value of
 $\mbox{BM}[r,i]$ and $\mbox{BM}[r,j]$ is $1$ and the other $0.$

We consider the highest $r$ that satisfies these conditions.

In a first step, if $\mbox{BM}[r,i]=1$ and $\mbox{BM}[r,j]=0$, then, as $i < j$
and as all columns has been sorted in increasing lexicographical
order, there must exist a row $r'$ higher than $r$ such that
$\mbox{BM}[r',i]=0$ and $\mbox{BM}[r',j]=1.$ We thus consider now
w.l.o.g that $\mbox{BM}[r,i]=0$ and $\mbox{BM}[r,j]=1.$

Among all pairs of indices $i$ and $j$ such that
$\mbox{BM}[r_X,i]=\mbox{BM}[r_X,j]=1$ and that there exits $r$ such
that $\mbox{BM}[r,i]=0$ and $\mbox{BM}[r,j]=1,$ let us consider one
pair $i'$ and $j'$, $1 \leq i'<j' \leq |V|$, that is associated to the
highest such $r$ that we denote $t.$

We now prove that $\mbox{BM}[t,\mbox{left}(X)]=0$ and
$\mbox{BM}[t,\mbox{right}(X)]=1.$ If $\mbox{BM}[t,\mbox{left}(X)]$ $=1$,
thus $i>\mbox{left}(X)$ and as $\mbox{BM}[t,i]=0$ and that the columns
are sorted in lexicographical order, there should exits an higher row
$r'$ such that $\mbox{BM}[r',\mbox{left}(X)]=0$ and
$\mbox{BM}[r',i]=1$, which contradicts $t$ to be the
highest such row. Thus $\mbox{BM}[t,\mbox{left}(X)]=0.$ Symmetrically,
the same argument holds to prove that
$\mbox{BM}[t,\mbox{right}(X)]=1.$
\end{preuve}

\begin{lemma}
 Let $X\in {\cal F}.$ Then $\mbox{Max}(X)\not=\emptyset$ if and only if
 there exists a row $t$ in $BM$ such that
 $\mbox{BM}[t,\mbox{left}(X)]=0$ and $\mbox{BM}[t,\mbox{right}(X)]=1$
 corresponding to a set $Y \in {\cal F}$ verifying $|Y| \geq |X|$. 
\label{uplemma}
\end{lemma}
\begin{preuve}
$(\Leftarrow)$ If a set $Y$ corresponds to a row $t$ in $BM$ such
  that $\mbox{BM}[t,\mbox{left}(X)]=0$ and
  $\mbox{BM}[t,\mbox{right}(X)]=1,$ $Y$ obviously overlaps $X$. As
  $|Y| \geq |X|,$ $\mbox{Max}(X)\not=\emptyset.$  $(\Rightarrow)$ Let us
  assume that $\mbox{Max}(X)\not=\emptyset$ and let $r_M$ be its row in
  $BM.$ Then, by lemma \ref{goodlemma}, there exists a row $t$ in $BM$
  such that $\mbox{BM}[t,\mbox{left}(X)]=0$ and
  $\mbox{BM}[t,\mbox{right}(X)]=1$ and such that $t$ is higher than or
  equal to $r_M$. As $\mbox{Max}(X)$ verifies $|\mbox{Max}(X)| \geq
  |X|,$ the set $Y$ corresponding to $r_M$ is also such that $|Y| \geq
  |X|.$
\end{preuve}

\begin{lemma}[\cite{Dahlhaus00}]
Let $X\in {\cal F}$ such that $\mbox{Max}(X)\not=\emptyset.$ Then
$\mbox{Max}(X)$ corresponds to the highest row $t$ in $BM$ such that
$\mbox{BM}[t,\mbox{left}(X)]=0$ and $\mbox{BM}[t,$ $\mbox{right}(X)]=1.$
\label{lemup}
\end{lemma}
\vspace*{-3mm}
[Notice that this row might be lower than the row corresponding to
  $X$. This is the case for $X_8$ and $X_{10}$ since
  $\mbox{Max}(X_{10})=X_8$ but also $\mbox{Max}(X_8)=X_{10}.$ in our example.]\\[3mm]
\begin{preuve}
 Let us assume that $\mbox{Max}(X)\not=\emptyset$ and let $r_M$ be its
 row in $BM.$ Then, by lemma \ref{goodlemma}, there exists a row $t$
 in $BM$ such that $\mbox{BM}[t,\mbox{left}(X)]=0$ and
 $\mbox{BM}[t,\mbox{right}(X)]=1$ and such that $t$ is higher than or
 equal to $r_M$. However, as such a row $t$ corresponds to a set
 overlapping $X$ and that $\mbox{Max}(X)$ is the largest of those
 sets in $LF$ order, $t=r_M$.
\end{preuve}

For example, in Figure \ref{matrix}, $\mbox{Max}(X_1) = X_9$ since
$\mbox{left}(X_1) = 1,$ $\mbox{right}(X_1) = 6$ and $X_9$ (row 2)
corresponds to the highest row with $0$ on the first column and $1$ on
the $6^{\mbox{th}}$.

Dahlhaus's approach for computing all $\mbox{Max}(X)$ is to identify for
each row $r$ corresponding to $X$ the highest row $t$ such that
$\mbox{BM}[t,\mbox{left}(X)]=0$ and $\mbox{BM}[t,\mbox{right}(X)]$ $=1.$
To do it efficiently, Dahlhaus reduces the problem to LCA
computations. We explain this reduction in the next section
\ref{LCA}. We then present another approach using class partitions
in \ref{partitions}. This new approach is much simpler to
implement than the LCA algorithm in its real linear worst case
complexity. Moreover, it allows an easy computation of the
lexicographical order of the columns.

\subsection{Computing all $\mbox{Max}(X)$ using LCA}
\label{LCA}

\begin{figure}[htb]
\centering \includegraphics[width=10cm]{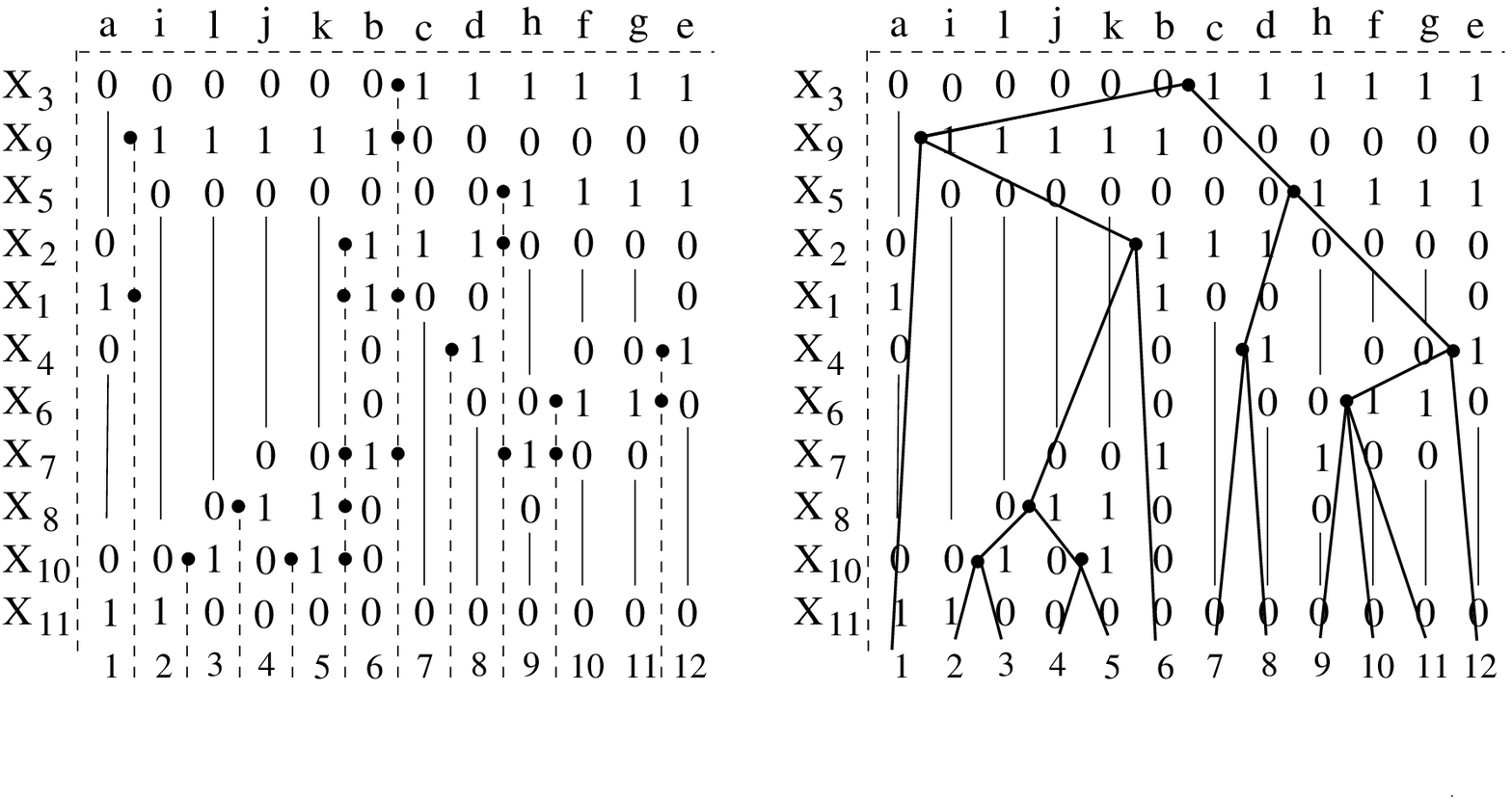}
\vspace{-0.7cm}
\caption{Example continued: Dahlhaus's tree built over a $BM$ matrix.}
 \label{lcapic}
\end{figure}

Let us consider all intermediate columns between all pairs
of columns in $\mbox{BM}.$ In those columns, for each row, we place a
point $\bullet$ between each motif $01$ or $10$. This is shown in Figure
\ref{lcapic} (left).  We link the highest point in each intermediate
column, if it exist, in a Dahlhaus's tree ($\mbox{DT}$) the following way: 
\begin{enumerate}
\item the root of the tree is the highest point. There can be only one
  root and there must be one root if one of the set $X \in {\cal F}$ differs
  from $V$. We assume this below;
\item we recurse the following process: each new point $np$ in the
  tree (root included) splits the submatrice in two subparts according
  to the intermediate column it is placed in; the left (resp. right)
  child of $np$ is the highest point in the left (right) part, if it
  exits. Note that the lexicographical order of the columns of $BM$
  insures that there can be at most one highest point in each part;
\item when a subpart does not contain any new point, a leaf per
$\mbox{BM}$ column in this subpart is created and attached as child to
the point that created the subpart. If this point is placed to the
left (resp. right) of this column, the child is a right (resp. left)
child. Each leaf is numbered with the number of the corresponding
column in $\mbox{BM}.$
\end{enumerate}

\noindent
An instance of such a tree is given in Figure \ref{lcapic} (right).

\begin{proposition}[\cite{Dahlhaus00}]
Let $X\in {\cal F}.$ Let $Y \in {\cal F}$ be the set corresponding to the
row of $LCA(\mbox{left}(X),$ $\mbox{right}(X))$ in $\mbox{BM}.$ If
$|Y|\geq |X|$, then $Y=\mbox{Max}(X).$ Otherwise $\mbox{Max}(X)=\emptyset.$
\end{proposition}
\begin{preuve}
Let $r$ be the number of the row of $LCA(\mbox{left}(X),$
$\mbox{right}(X))$ in $\mbox{BM}$ and let $l$ be the position of the
column in $BM$ that is just before the point representing
$LCA(\mbox{left}(X),$ $\mbox{right}(X)).$

First, $BM[r,l]=0$ and $BM[r,l+1]=1.$ Suppose a contrario that
$BM[r,l]=1$ and $BM[r,l+1]=0.$ As all columns of $BM$ are sorted in
lexicographical order, there must exists an higher row $r'$ such that
$BM[r,l]=0$ and $BM[r,l+1]=1.$ and thus a point in the intermediate
column between $l$ an $l+1$ higher than that in row $r$, which
contradicts the construction of $DT.$

We now prove that $BM[r,\mbox{left}(X)]=0$ and
$BM[r,\mbox{right}(X)]=1.$ A contrario, suppose that
$BM[r,\mbox{left}(X)]=1.$ Then, again, as the columns of $BM$ are
sorted in lexicographical order, there must exists an higher row $r'$
such that $BM[r',\mbox{left}(X)]=0$ and $BM[r',l]=1.$ This again
contradicts the construction of $DT.$ A similar argument holds for the
right side.

We then prove that $r$ is the highest row with this property. Assume a
contrario that there exist an higher row $r'$ such that
$BM[r',\mbox{left}(X)]=0$ and $BM[r',\mbox{right}(X)]=1.$ Then there
would have been a split $01$ somewhere in this row that would have
separated $\mbox{left}(X)$ and $\mbox{right}(X).$ This implies that
there would have been a node in $DT$ in a row higher than or equal to
$r'$ that would have split $\mbox{left}(X)$ and $\mbox{right}(X),$
which contradicts $r$ to be the number of the row of
$LCA(\mbox{left}(X),$ $\mbox{right}(X)).$

If $|Y|\geq |X|$, by Lemma \ref{lemup} $\mbox{Max}(X)\not=\emptyset$ and
the set $Y$ that corresponds to $r$ is such that $Y=\mbox{Max}(X).$

If $|Y|<|X|$, since no row $r'$ higher than $r$ can verify
$BM[r',\mbox{left}(X)]=0$ and $BM[r',\mbox{right}(X)]=1,$ by Lemma
\ref{uplemma} $\mbox{Max}(X)\not=\emptyset.$
\end{preuve}

\noindent
For example, $X_9$ corresponds to the row of $LCA(1,2)=LCA(\mbox{left}(X_{11}),$ $\mbox{right}(X_{11})).$ As
$|X_9|\geq |X|,$ $X_9=$ $\mbox{Max}(X_{11}).$


\subsection{Computing all $\mbox{Max}(X)$ using set partitioning}
\label{partitions}

We present below an alternative approach that permits avoiding LCA
queries. Moreover, the lexicographical column order appears 
as a by-product.

We manipulate sorted partitions of $V$ that we refine by each $X \in
{\cal F}$ taken in $\mbox{LF}$ order, that is, in decreasing order of
their sizes. The initial partition is the whole set $V$ and denoted
$P_V$.  For clarity, a set in a partition is called a {\em part}. In
each partition the order of the parts is important, but the order of
elements in a same part is not. Let $C=\{v_1,\ldots,v_k\}$ be a part
in a partition. Refining $C$ by $X \in {\cal F}$ consists in
extracting all $v_i \in X$ in $C$ and create a new part $C''$ with all
those $v_i$. The remaining $v_i \not\in X$ in $C$ form a new part $C'$
and $C$ is replaced in the current partition by $C'C''$. If $C$ only
contains elements of $X$ as well as if it contains none, $C$ remains
unchanged in the partition. Refining a partition $P$ by a set $X \in
{\cal F}$ consists in refining successively all parts in $P$. We note
this refinement $P|_X.$

For example (continued), if
$P=\{a\}\{i,j,k,l\}\{b\}\{c,d\}\{e,f,g,h\}$ and $X = X_4 =
\{d,e\}$, $P|_X = \{a\}\{i,j,k,l\}\{b\}\{c\}\{d\}\{f,g,h\}\{e\}.$\\

\noindent
Our approach requires 3 steps:
\begin{enumerate}
\item refine $P_V$ by all $X \in {\cal F}$ taken in
  $\mbox{LF}$ order;
\item then compute for each $X \in {\cal F}$ the values of
  $\mbox{left}(X)$ and $\mbox{right}(X)$ and sort all $X \in {\cal
  F}$ in a special order in regard with these values;
\item eventually refine $P_V$ again by all $X \in {\cal F}$
taken in $\mbox{LF}$ order but using the informations computed in step
2 to compute all $\mbox{Max}(X).$
\end{enumerate}
\noindent
We detail below each step. 

\paragraph{\bf \em Step 1 \-- Refining $P_V.$}

Let us consider the final partition we obtain after refining $P_V$ by
each $X \in {\cal F}$ taken in $LF$ order. We note this partition
$P_f$.

\begin{lemma}
The elements of $P_f$ are sorted accordingly to the lexicographical order
of the columns of $BM.$
\label{lexico}
\end{lemma}
\begin{preuve}
Refining a partition consists in lexicographically sorting a row of
$BM$ touching only the $1$ in the row but also keeping the global
order already defined by the sets in the partition. Thus refining
partitions from $P_V$ in $LF$ order consists in lexicographically
ordering $BM$ from the top row to the bottom.
\end{preuve}

\noindent
For example (continued), on the data in Figure \ref{allexample},
  $P_f=\{a\}\{i\}\{l\}\{j\}\{k\}$$\{b\}$ $\{c\}$$\{d\}\{h\}\{f,g\}\{e\}.$
  Note that equal columns of $BM$ are in the same part of $P_f$ on
  which we fix an arbitrary order.

\paragraph{\bf \em Step 2 \-- Computing all $\mbox{left}(X)$ and $\mbox{right}(X)$ values.}

We then compute all $\mbox{left}(X)$ and $\mbox{right}(X)$ values on
$P_f.$ This can be done easily in $O(|{\cal F}|+n)$ time by scanning each
$X \in {\cal F}$ and keeping the minimum and maximum position of one
of its element in $P_f$. We also compute a data structure $AM$ that
for each position $1 \leq i \leq |V|$ of $P_f$ gives a list of all $X
\in {\cal F}$ such that $i=\mbox{right}(X)$. All those lists are sorted in
increasing order of $\mbox{left}(X).$ The structure also allows an
element $X \in {\cal F}$ to be removed from the list
$AM[\mbox{right}(X)]$ in $O(1)$ time. This can be insured for instance
using doubly linked list to implement each list, and the whole
structure can easily be built in $O(n+m)$ time using bucket sorting.

\paragraph{\bf \em Step 3 \-- Refining $P_V$ again and identifying all $\mbox{Max}(X).$}

The main idea is the following. Assume that at a step of the
refinement process in $LF$ order we refine a part $C=\{v_1,\ldots,
v_k\}$ of a partition $P$ by $Y\in {\cal F}$ and that it results two
non empty parts $C'C''.$

\begin{lemma} 
Let $X \in {\cal F}$ such that $|X| \leq |Y|$,  $\mbox{left}(X) \in C'$
and $\mbox{right}(X)\in C''.$ Then $Y=\mbox{Max}(X).$
\label{assignmax}
\end{lemma}
[Note that if $|X| = |Y|$ then $X$ could be before $Y$ in $LF$ order.]\\
\begin{preuve}
Let $r$ be the row corresponding to $Y$ in $BM$. As $\mbox{left}(X)\in
C'$ and $\mbox{right}(X) \in C''$, then $BM[r,\mbox{left}(X)]=0$ and
$BM[r,\mbox{right}(X)]=1,$ and $Y$ obviously overlaps $X.$ As $|X|
\leq |Y|,$ $\mbox{Max}(X)\not=\emptyset.$ Moreover, the row $r$ is the
highest such that $BM[r,\mbox{left}(X)]=0$ and
$BM[r,\mbox{right}(X)]=1$ since otherwise the elements of $X$ would
have been split by a set bigger that $Y$ in the $LF$ order. Thus, by
Lemma \ref{lemup}, $Y=\mbox{Max}(X).$
\end{preuve}

\noindent
The last phase of the algorithm thus consists in refining $P_V$ again by
all $Y \in {\cal F}$ taken in $LF$ order. We first initialize all
values $\mbox{Max}(X)$ to $\emptyset$. Each time a new split $C'C''$
appears (say between positions $l$ and $l+1$), for all $v \in C''$ all
lists $AM[v]$ are inspected the following way: let $X$ be the top of
one of those the list; while $\mbox{left}(X)\leq l$, $X$ is popped off
the list and $\mbox{Max}(X) \leftarrow Y$. After having refined with
$Y$, if there is no more $Y' <_{LF} Y$ such that $|Y'|=|Y|$, all sets
of the same size than $Y$ are removed from the $AM$ structure.\\

\begin{lemma}
Our algorithm correctly computes in 3 steps all $\mbox{Max}(X), \; X
\in {\cal F}$.
\label{correctness}
\end{lemma}
\begin{preuve}
In step 1 the lexicographical order of the columns of $BM$ is computed
as a partition $P_f$ (Lemma \ref{lexico}). In step 2 all values
$\mbox{left}(X)$ and $\mbox{right}(X)$, $X \in {\cal F},$ are computed
and the $AM$ structure is built.  In step 3, the correctness of the
computation relies on the following observation: for each new
partition $P$ created after a refinement, all sets $X$ remaining in
$AM$ are such that $\mbox{left}(X)$ and $\mbox{right}(X)$ belong to
the same part in $P$. This is obviously true since otherwise they would
have been split by a previous refinement and removed of $AM$. This has
for consequence that after a split of a set $C$ in $C'C''$ by a set
$Y$, testing if $\mbox{left}(X) \in C'' $ and $\mbox{right}(X) \in
C''$ for all sets in $AM$ is equivalent to test if $\mbox{right}(X)
\in C''$ and $\mbox{left}(X)\leq l$, where $l$ is the left position in
$P$ of the split between $C$ and $C''$. Moreover, as each set taken in
$LF$ order and used for a possible refinement is removed of $AM$ after
having processed all the sets of the same size, when a set $Y$ splits
a part $C$ in $CC''$, all sets in $AM$ are such that $|X|\leq |Y|$. We
thus fulfill all requirements of Lemma \ref{assignmax} and
$Y=\mbox{Max}(X).$ Thus, if a value $\mbox{Max}(X)$ is assigned by our
algorithm, it is assigned with the right one.

Now, suppose that a set $X$ admits a set $Y$ as $\mbox{Max}(X)$. It is
guaranteed that a certain step of the algorithm $Y$ has been assigned
to $\mbox{Max}(X)$ since that by definition $|X|\leq |Y|$ which
implies that $X$ is still in $AM$ when $Y$ is processed and that by Lemma
\ref{lemup} $\mbox{left}(X)\not\in Y$ and $\mbox{right}(X)\in Y.$ The
set $Y$ has thus split a part $C$ in a partition in $C'C''$ such that
$\mbox{right}(X)>l$ and $\mbox{left}(X)\leq l$ where $l$ is the left
position in $P$ of the split between $C$ and $C''.$
\end{preuve}

\begin{figure}[htb]
\centering \includegraphics[width=8cm]{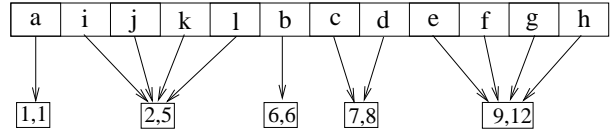}
\caption{Example continued: implementation of
$P=\{a\}\{i,j,k,l\}\{b\}\{c,d\}\{e,f,g,h\}.$}
 \label{lcapic2}
\end{figure}

\noindent
It remains to explain how a partition refinement can be efficiently
implemented. We exploit the fact that element's order inside each part
of a partition has no importance to obtain a simple implementation: a
partition is represented as a table of size $n$ in which each cell
contains (a) an element of $V$ and (b) a pointer to the part of the
partition in which it is contained. A part is
represented by a pair of its bounds on this table. Figure
\ref{lcapic2} shows such an implementation.

\noindent 
Refining a partition $P$ by a set $Y$ can be done in $O(|Y|)$ the
following way. Let $[i,j]$ be the bounds of a part $C$ such that $C
\not\subset Y$ (easily testable). Let $k$ be the number of elements of
$Y$ that belongs to the subtable $[i,j]$, $ 1 \leq k \leq j-i$. We
swap elements in the subtable $[i,j]$ to place all $k$ elements
belonging to $Y$ at the end of this subtable. We then adjust the
bounds of $C$ to $[i,j-k]$ and create a new set $[j-k+1,j]$ on which
the $k$ elements of $Y$ now point.

\begin{theorem}
The identification of all $\mbox{Max}(X), \; X \in {\cal F},$ using
partition refinement can be done in $\Theta(n+|{\cal F}|)$ time. 
\end{theorem}
\begin{preuve}
By Lemma \ref{correctness} the algorithm is correct. Steps 1 and 2 are
$\Theta(|{\cal F}|+n)$ time. In step 3, the fact that all lists in
$AM$ are sorted in increasing order of $left()$ values insures that
when a set $Y$ splits a part $C$ in $C'C''$, identifying and popping
off all sets $X$ such that $\mbox{left}(X) \in C$ and $\mbox{right}(X)
\in C''$ can be done in $\Theta(|C|+K+1)$ time, where $K$ is the number
of such sets. Removing a set out of $AM$ is $O(1)$ time, thus the
total of time managing $AM$ is $\Theta(|{\cal F}|+n)$ time.
\end{preuve}

The whole algorithm has been implemented in its real worst case time
complexity and is freely available in \cite{OurImpl07}.

\section{Computing a subgraph of the overlap graph}

In some applications like in \cite{McConnell04} it is useful to get a
spanning tree of all overlap classes of $OG({\cal F},E).$ The approach
of \cite{McConnell04} is to first compute Dahlhaus's graph and then
compute spanning trees of the connected components of the overlap
graph using a quite complex add-on. We thus explain in this section
how to simply modify Dahlhaus's approach to compute a subgraph of the
overlap graph instead of $D({\cal F},L).$ The size of the subgraph is
linear but it has the same connected components than the overlap graph
and it is thus easy from it to compute spanning trees of the overlap
graph. The idea of the modification is the following.

\begin{lemma}
Let $X,Y \in {\cal F}$ such that $X \cap Y \not= \emptyset$, such that
$\mbox{Max}(X)\not=\emptyset$ and such that $|X| \leq |Y| \leq
|\mbox{Max}(X)|.$ Let $r_Y$ be the row of $Y$ in $BM.$ If
$BM[r_Y,\mbox{left}(X)]=0,$ $Y$ overlaps $X.$ Otherwise, (a) if
$BM[r_Y,\mbox{right}(X)]=0$, then $Y$ overlaps $X$, and (b) if
$BM[r_Y,\mbox{right}(X)]=1$, then $Y$ overlaps $\mbox{Max}(X).$
\label{quintuple}
\end{lemma}
\begin{preuve}
Let $r_X$ be the row of $X$ in $BM$, and $r$ that of $\mbox{Max}(X).$
If $BM[r_Y,left(X)]=0,$ as $BM[r_X,left(X)]=1$, that $X \cap Y \not=
\emptyset$ and that $|X| \leq |Y|$, $Y$ overlaps $X$. Assume now that
$BM[r_Y,left(X)]=1.$ Case (a): if $BM[r_Y,\mbox{right}(X)]=0,$ then,
as $BM[r_X,right(X)]=1$, with the same arguments that above $Y$
overlaps $X.$ Case (b): if $BM[r_Y,\mbox{right}(X)]=1,$ then, as by
Lemma \ref{lemup}, $BM[r,\mbox{right}(X)]=1$ and
$BM[r,\mbox{left}(X)]=0$, and that $|Y| \leq |\mbox{Max}(X)|$, $Y$
overlaps $\mbox{Max}(X)$.
\end{preuve}

\begin{figure}[htb]
\vspace{-0.3cm}
  \centering
\includegraphics[width=9cm]{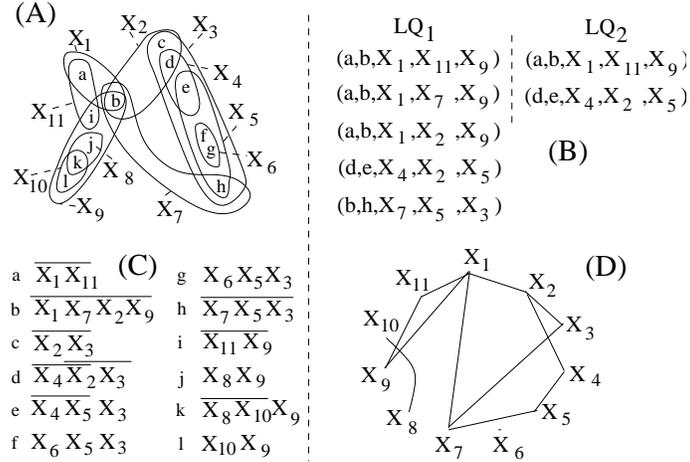}
\caption{Global example (continued): (A) input family of 11 sets; (B)
  $\mbox{LQ}_1$ and $\mbox{LQ}_2$ lists in which $\mbox{right}(X)$ and
  $\mbox{left}(X)$ heve been replaced by $Pf_{\tiny \mbox{right}(X)}$
  and $Pf_{\tiny \mbox{left}(X)}$ ; (C) $SL$ lists; (D) the resulting subgraph of
  the overlap graph.}
 \label{allexample2}
\vspace{-0.3cm}
\end{figure}

We modify the construction of Dahlhaus's graph the following way. We
still consider intervals $X..YW..$ on $SL(v)$ lists such that
$\mbox{Max}(X)\not=\emptyset$ and $|W|\leq |\mbox{Max}(X)|$, but instead
of creating a chain $X-..Y-W-..$ in $D({\cal F},L)$, we create an edge
$(X,\mbox{Max}(X))$ (if it does not already exists) and a list of
quintuples $(\mbox{left(X)},\mbox{right}(X),X,Y,\mbox{Max}(X)),$
$(\mbox{left(X)},\mbox{right}(X),X,W,\mbox{Max}(X)),..$ for all the
elements in the interval distinct of $X$ and $\mbox{Max(X)}$. All
quintuples for all intervals are placed in the same list $LQ_1$. Note
that if an element belongs to 2 intervals, a unique quintuple is
formed with the rightest interval.

To apply Lemma \ref{quintuple}, if suffices for each 
$(\mbox{left(X)},\mbox{right}(X),X,Y,\mbox{Max}(X))$  to test if $Y$ belongs to
$SL(Pf_{\mbox{\tiny left(X)}})$. If not, we then create an edge $(X,Y).$
Otherwise, we test if $Y$ belongs to $SL(Pf_{\mbox{\tiny right}(X)}).$ If
not, we also create an edge $(X,Y).$ However, if it does, we create an
edge $(Y,\mbox{Max}(X)).$

For complexity issues we need to perform those tests at a glance for
all quintuples in $LQ_1$. We do it in two phases. In the first phase
we search for all $Y$ in $SL(Pf_{\tiny \mbox{left(X)}})$. If $Y$ does not
belong to $SL(Pf_{\mbox{\tiny left(X)}})$, we add the quintuplet
$(\mbox{left(X)},\mbox{right}(X),X,Y,\mbox{Max}(X))$ to a second list
$LQ_2$. In the second phase, if $LQ_2$ is not empty, for all
 $(\mbox{left(X)},\mbox{right}(X),X,Y,\mbox{Max}(X))$ in
$LQ_2$ we search $Y$ in $SL(Pf_{\mbox{\tiny right(X)}}).$

We assume below that all $SL(v)$ lists are sorted accordingly to the
$LF$ order instead of being simply sorted by increasing sizes. To
efficiently compare $LQ_1$ with all $SL(v)$ lists it suffices to sort
the list $LQ_1$ accordingly to $\mbox{left}(X)$ and then sort all
quintuples with the same $\mbox{left}(X)$ value in the $LF$ order of
$Y.$ This can be done in $O(n+|{\cal F}|)$ time using bucket
sorting. The comparison of $LQ_1$ and the tables $SL()$ can then be
done in $O(n+|{\cal F}|)$ time by comparing simutaneously $|V|$ sorted
lists.  The same approach holds for $LQ_2.$ We thus have:

\begin{theorem}
A subgraph of the overlap graph of ${\cal F}$ having the same
connected components can be computed in $O(n+|{\cal F}|)$ time.
\end{theorem}
\begin{preuve}
Lemma \ref{quintuple} insures that the new graph is a subgraph of the
overlap graph. To prove that they have the same connected component,
it thus suffices to prove that if two sets $A$ and $B$ overlap, there
exists a path connecting $A$ and $B$ in the subgraph. The following
observation is the base of the proof: let $X..Y..Z$ sorted by
increasing size on the same $SL(v)$ and such that $|Y|\leq
|\mbox{Max}(X)|$, $|\mbox{Max}(X)| \leq |\mbox{Max}(Y)|$ and $|Z|\leq
|\mbox{Max}(Y)|.$ Then there exists a path between all sets $X,Y,Z$ in
the new subgraph since by construction $X$ and $\mbox{Max}(X)$ are
connected, $Y$ is connected to $X$ or $\mbox{Max}(X)$, $Y$ is
connected to $\mbox{Max}(Y)$ and eventually $Z$ is connected to $Y$ or
$\mbox{Max}(Y)$.

Now let $v \in A \cap B.$ Assume w.l.o.g. that $|A|\leq |B|.$ Then
$\mbox{Max}(A)\not=\emptyset$ and $|\mbox{Max}(A)| \geq |B|.$ Therefore,
in $SL(v)$, there exits a series (potentially empty) of $k$ sets
$A..Y_1..Y_2..Y_k..B$ such that $|B| \leq |\mbox{Max}(Y_k)|,$ $|Y_k|
\leq |\mbox{Max}(Y_{k-1})|,$ and $|Y_1| \leq |\mbox{Max}(A)|.$ By
induction on the series using the previous observation there exits a
path from $A$ to $B$ in the subgraph.

The subgraph can obviouly been built in $O(n+|{\cal F}|)$ time since
all steps can be done in this time.
\end{preuve}

\noindent
An example (continued) of the resulting subgraph is shown in Figure
\ref{allexample2}.

\end{document}